\documentstyle[12pt]{article}
\begin{document}
\input epsf.tex
\def\a{\alpha}
\def\b{\beta}
\def\ch{\chi}
\def\d{\delta}
\def\e{\epsilon}
\def\f{\phi}
\def\g{\gamma}
\def\h{\eta}
\def\i{\iota}
\def\j{\psi}
\def\k{\kappa}
\def\l{\lambda}
\def\m{\mu}
\def\n{\nu}
\def\o{\omega}
\def\p{\pi}
\def\q{\theta}
\def\r{\rho}
\def\s{\sigma}
\def\t{\tau}
\def\u{\upsilon}
\def\x{\xi}
\def\z{\zeta}
\def\D{\Delta}
\def\F{\Phi}
\def\G{\Gamma}
\def\J{\Psi}
\def\L{\Lambda}
\def\O{\Omega}
\def\P{\Pi}
\def\S{\Sigma}
\def\U{\Upsilon}
\def\T{\Theta}

\def\Ab{\bar{A}}
\def\gi{g^{-1}}
\def\li{{ 1 \over \l } }
\def\lb{\l^{*}}
\def\zb{\bar{z}}
\def\ub{u^{*}}
\def\Tb{\bar{T}}
\def\pp {\partial }
\def\pb {\bar{\partial }}
\def\be{\begin{equation}}
\def\ee{\end{equation}}

\def\phib{\phi^\dagger}
\def\beq{\begin{eqnarray}}
\def\eeq{\end{eqnarray}}
\def\ben{\begin{eqnarray}}
\def\een{\end{eqnarray}}
\def\tr{{\rm Tr}}
\def\X{{\bf X}}
\def\sech{{\rm sech}}

\hsize=16.5truecm
\addtolength{\topmargin}{-0.6in}
\addtolength{\textheight}{1.5in}
\vsize=26truecm
\hoffset=-.6in
\baselineskip=7 mm
\thispagestyle{empty}
\rightline{hep-th/9907068}
\rightline{\today}
\vskip2cm
\centerline{\Large\bf Vortex String Dynamics }
\centerline{\Large\bf in an External Antisymmetric Tensor Field}

\vskip 1cm

\centerline{Kimyeong Lee\footnote{e-mail:
kimyeong@phya.snu.ac.kr}, Q-Han Park\footnote{e-mail:
qpark@nms.kyunghee.ac.kr} and H.J. Shin\footnote{e-mail:
hjshin@nms.kyunghee.ac.kr}}

\vskip2mm
\centerline{$^1$Physics Department and CTP, Seoul National University, Seoul 130-743,
Korea}
\vskip3mm
\centerline{$^{2,3}$Department of Physics and Research Institute of Basic
Science}
\centerline{Kyung Hee University, Seoul 130-701,  Korea}
\vskip 2cm
\centerline{\bf ABSTRACT}
\vskip 5mm

We study the Lund-Regge equation that governs the motion of strings in a constant background
antisymmetric tensor field by using the duality between the Lund-Regge
equation and the complex sine-Gordon equation. Similar to the cases of vortex filament
configurations in fluid dynamics, we find various exact solitonic string configurations
which are the analogue of the Kelvin wave, the Hasimoto soliton and the smoke ring.
In particular, using the duality relation, we obtain a completely new type of configuration
which corresponds to the breather of the complex sine-Gordon equation.
\vskip 5mm

\newpage

\setcounter{footnote}{0}

\vskip 5mm

\section{Introduction}

There has been some studies on the dynamics of a string coupled to
an antisymmetric tensor field. Such a coupling was first introduced
by Kalb and Ramond in the context of string theory \cite{kalb}. It
also appears in the study of the relativistic motion of vortex
strings in a superfluid by Lund and Regge \cite{lund}, where the antisymmetric
tensor field induces the Magnus force acting on the vortex string by
uniform fluid density. In particular, Lund and Regge has shown that
the dynamical equation for vortex strings can be `dualized'
to the complex sine-Gordon equation \cite{lund}, and they have also found a
soliton solution of the complex sine-Gordon equation.
More recently, using a simple ansatz, Zee found the smoke ring solution of
the Lund-Regge equation \cite{zee}.
However, despite the dualization of the Lund-Regge equation to the complex
sine-Gordon equation, which is exactly integrable with many known exact solutions,
nothing much is known about the exact solutions of the Lund-Regge equation itself.

In the context of fluid dynamics, we encounter vortex filaments in an
incompressible inviscid fluid \cite{saffman}. The relevant dynamics of vortex
filaments is described in terms of variables $\X(\t, \s)$, which represent the vortex
position where $\s$ measures the length of the vortex filament. When we neglect both
vortex mass and long range interaction, and also introduce a short-distance cutoff,
the dimensionless form of the dynamical equation for the vortex $X(\t, \s)$
becomes the Da Rios-Betchov equation \cite{darios}:
\begin{equation}
\dot\X = \X'\times \X''
\label{Bet}
\end{equation}
where the dot denotes $\partial/\partial \t$ and the prime denotes
$\partial/\partial \s$.
Various exact solutons of the Da Rios-Betchov equation are known
and their stability properties have been analyzed. Among many
solutions, the Kelvin wave, a helical configuration of vortex
filament, and the smoke ring solutions arise as simple cases.
Another interesting solution is the solitary kink wave which
propagates along a straight vortex filament line. This was first
found by Hasimoto  \cite{hasimoto} by mapping the Da Rios-Betchov
equation into the nonlinear Schr\"odinger (NLS) equation and
finding soliton solutions of the NLS equation. The stability of the
Kelvin wave has been also studied \cite{samuels}. There are also
more recent studies of this equation \cite{yasui}.

In the case of the Lund-Regge equation, the smoke ring solution found by Zee suggests
that other types of solitonic solutions might exist as well. Moreover, the duality between
the Lund-Regge and the complex sine-Gordon equations implies that exact solutions,
e.g. solitons \cite{lund}, of the complex sine-Gordon equation have counterparts in
the vortex system. So far, however, explicit expressions of these solutions have not been
known despite of the dualization of the Lund-Regge equation to the integrable complex sine-Gordon
equation. This is mainly due to the fact that the duality relation between two equations obtained by
Lund and Regge is incomplete. That is, their dualization is only one-directional: it
allows us to express the variables of the complex sine-Gordon equation in terms of those of
the Lund-Regge equation but not vice versa. Recently, the reverse direction has been found
by utilizing the associated linear equation of the complex sine-Gordon equation, in addition with an abstract
generalization of the Lund-Regge equation using group theory \cite{Park2}.

In this paper, we analyze the Lund-Regge model using its dualization to the complex
sine-Gordon model. The duality between these two models is explained in detail in terms of the
 $SU(2)/U(1)$-gauged Wess-Zumino-Novikov-Witten (WZNW) model with a potential term. This allows us to understand
symmetries of the Lund-Regge model from different perspectives. By making use of exact solutions
of the complex sine-Gordon equation, we find various vortex-type configurations of the Lund-Regge
equation similar to vortex filament configurations in fluid dynamics such as the Kelvin wave, the
smoke ring, and the Hasimoto soliton. In particular, we obtain the breather-type vortex solution by
dualizing the breather solution of the complex sine-Gordon equation.

The plan of this paper is as follows. In Sec. 2, we review the
Lund-Regge model and its relation to the complex sine-Gordon model.
In Sec. 3, we explain about the Wess-Zumino-Novikov-Witten sigma model formulation of the complex sine-Gordon
model, and derive exact solutions of the complex sine-Gordon
equation using the B\"{a}cklund transformation.
In Sec. 4, we dualize these solutions to the
configurations of vortex strings. We obtain the smoke ring, Kelvin
wave, Hasimoto soliton, and the breather. We conclude with some
remarks in Sec. 5.

\section{The Lund-Regge Model}

The dynamics of a charged particle in an external electromagnetic field is determined by a
Lorentz force acting on a particle. From the view point of a particle Lagrangian, this is done
by coupling the vector gauge potential of electromagnetism to a charged particle in a gauge invariant
way. In the case of string dynamics in an external field, the antisymmetric tensor field $B_{\m \n}$
plays the role of a gauge connection for string and the corresponding field strength ($H=dB$) induces
the Magnus force acting on string. In physical applications, the antisymmetric tensor field is not a
fundamental physical variable such as the vector gauge potential but appears as an effective description
of external fields. For example, the vortex string in an idealized superconductor in zero temperature
is described by a Maxwell-Higgs system with a uniform external electric charge density \cite{kimyeong}.
In this case, the Magnus force is generated by the electric charge of the Higgs field which neutralizes
the background charge. In the following, we will not concern about any specific physical applications.
Instead, we regard the antisymmetric tensor field simply as an external field appearing in the context
of a string theory.

The Lagrangian for the string coupled to an antisymmetric tensor field is given by
\begin{equation}
{\cal L}_{\rm string}  = -{\mu \over 2} \partial^\alpha X^\mu
\partial_\alpha X_\mu   + B_{\mu\nu} \epsilon^{\alpha\beta} \pp_\a
X^\mu \pp_\b X^\nu,
\end{equation}
where $ ( \tau ,\sigma )$ is the string world-sheet coordinates and
$\alpha = \tau , \sigma ; ~ \mu =0,1,2,3. $ The string tension is
given by $\mu$. Here, we only consider the closed string or the
infinite size string without any ends. This Lagrangian is invariant
under the conformal transformation $\tau\pm \sigma \rightarrow
f_\pm (\tau \pm\sigma) $ where $f_\pm$ are arbitrary functions. The
conformal symmetry results in a constraint as the vanishing of the
two dimensional energy-momentum tensor,
\begin{eqnarray}
 &\ \dot{X}^\mu \dot{X}_\mu +\ X'^\mu X'_\mu = 0, \nonumber \\
 &\ \dot{X}^\mu X'_\mu = 0.
\label{emtensor}
\end{eqnarray}
On the other hand, the classical equation for the string coming from the variation on $X^\m$ is
\begin{equation}
 \mu \partial_\alpha^2 X^\mu = H^{\mu\nu\rho} \epsilon^{\alpha\beta}
\partial_\alpha X_\nu \partial_\beta X_\rho,
\end{equation}
where $H_{\mu\nu\rho}\equiv  \pp_\m B_{\n \r} + \pp_\n B_{\r \m}
+ \pp_\r B_{\m \n} $ is the gauge invariant field strength of
$B_{\m\nu}$.
The  Magnus force acting on the string is a stringy generalization of
the Lorentz force, and appears in the literature in many disguise. In
particular, Lund and Regge have shown that the relativistic motion of
vortices in a uniform static field  $H$ is described precisely in this
way; in a Lorentz frame in which $X^0 =\t$, the system is governed by
the gauge fixed equation of motion,
\be
(\pp_{\t}^{2}-\pp_{\s}^{2})X^{i} + c
\e_{ijk}\pp_{\t}X^{j}\pp_{\s}X^{k}=0,  ~ ~ i =1,2,3
 \label{vortex}
\ee
where $H_{ijk} = c\mu \epsilon_{ijk}$,
and also by the quadratic constraints:
\be
(\pp_{\t}X^{i})^2 +(\pp_{\s}X^{i})^2=1,
~~(\pp_{\s}X^{i})(\pp_{\t}X^{i})=0.
\label{constr}
\ee
Here, $X^{i }(\s , \t ); i =1,2,3$ are the vortex coordinates and $\s , ~ \t$ are local
coordinates on the string world-sheet. In the no-coupling limit ($c=0$), this equation describes the
transverse modes of the 4-dimensional Nambu-Goto string in the orthonormal gauge. The critical step
leading to the integration of the vortex equation (\ref{vortex}) and (\ref{constr}) is to interpret
the equation as the Gauss-Codazzi integrability condition for the embedding of a surface, i.e. the
embedding of the string world-sheet projected down to the $X^{0}=\t $ hypersurface into the
3-dimensional Euclidean space, $X^{0} = \t $. The induced metric on the projected world-sheet is given by
\be
ds^2 = ( \pp_{\s }\vec{X})^2 d\s^2 +
2 ( \pp_{\s }\vec{X} \cdot \pp_{\t }\vec{X} ) d\s d\t +
( \pp_{\t }\vec{X})^2 d\t^2 ,
\ee
or
\be
ds^2 =\cos^{2}\phi d\s^2 + \sin^{2}\phi d\t^2 ,
\ee
by parameterizing $( \pp_{\s }\vec{X})^2 =\cos^{2}\phi ,~ ( \pp_{\tau }\vec{X})^2 =\sin^{2}\phi $
according to Eq. (\ref{constr}). The unit tangent vectors, $\vec{N}_{1}$ and $ \vec{N}_{2}$, spanning
the plane tangent to the surface, and the unit normal vector $\vec{N}_{3}$ consisting a moving frame
are given by
\be
\vec{N}_{1} = {1 \over |\pp_{\s}\vec{X}|}\pp_{\s}\vec{X}
,
~~~
\vec{N}_{2} = {1 \over |\pp_{\t}\vec{X}| }\pp_{\t}\vec{X},
~~~
\vec{N}_{3}={1 \over | \pp_{\s }\vec{X} \times \pp_{\t }\vec{X} | }
 \pp_{\s }\vec{X} \times \pp_{\t }\vec{X} .
\ee
The vectors $(\vec{N}_{i}; ~ i=1,2,3)$, given coordinates $u_{1}, u_{2}$
on the surface, satisfy the equation of Gauss and Weingarten (with
$a=1,2$):
\be
{\pp \vec{N}_{a} \over \pp u_{b}} = \G^{c}_{ab}\vec{N}_{c} +
L_{ab}\vec{N}_{3}, ~~~
{\pp \vec{N}_{3} \over \pp u_{a}} = -g^{ab}L_{bc}\vec{N}_{c} ,
\ee
where $\G^{c}_{ab}$ are the Christoffel symbols and the $L_{ab}$ are the components of the extrinsic
curvature tensor. They are a set of overdetermined linear equations and the consistency of which requires
the Gauss-Codazzi equation:
\be
R_{abcd}=L_{ac}L_{bd}-L_{ad}L_{bc} , ~~~ L_{ab;c}=L_{ac;b},
\label{Gauss}
\ee
where the semicolon denotes covariant differentiation on the surface and $R_{abcd}$ are the components
of its Riemann tensor. From Eq. (\ref{Gauss}), it follows that there exists a field $\eta $ such that
\be
L_{12}=\cot \phi {\pp \eta \over \pp u_{2}} , ~~~
{1 \over 2}(L_{11} + L_{22}) = \cot \phi {\pp \eta \over \pp u_{1}} .
\ee
We introduce the light-cone coordinates $z=(\s + \t )/2 , ~~ \zb =(\s -\t )/2 $ and make the coordinate
transformation: $z \rightarrow z/\l , ~~ \zb \rightarrow \l \zb$, under which the Gauss-Codazzi equation
is invariant due to its Lorentz invariance. In this case, the Gauss-Weingarten equation in the spin-1/2
representation becomes the linear equation of the inverse scattering \cite{lund2}:
\be
\pp \Phi = -(U_{0} + \l U_{1})\Phi , ~~~
\pb \Phi = -(V_{0} + \l^{-1} V_{1})\Phi ,
\label{Lax}
\ee
where
\ben
U_{0}+\l U_{1} &=& - \pmatrix{ i c  \l /4 + i \pp \eta \cos 2\phi   / 2 \sin^2 \phi
 & - \pp \phi + i \pp \eta \cot \phi \cr
\pp \phi + i \pp \eta \cot \phi & -i c \l /4 -i  \pp \eta\cos 2\phi  / 2\sin^2 \phi },
\nonumber \\
V_{0}+\l^{-1} V_{-1} &=& -{i \over 4}\pmatrix{ -c \cos 2\phi /\l - 2 \pb \eta /\sin^2 \phi
&  -c  \sin 2\phi / \l  \cr
-c  \sin 2\phi /\l & c \cos 2\phi /\l + 2 \pb \eta /\sin^2 \phi }.
\label{lin}
\een
The integrability condition of the overdetermined linear equation in Eq. {\ref{Lax}),
\be
[\pp +U_{0}+\l U_{1},~ ~  \pb +V_{0}+\l^{-1} V_{-1} ]=0,
\label{int}
\ee
then becomes the complex sine-Gordon equation:
\ben
\pp \pb \phi - {c^2 \over 2} \sin 2\phi + {\cos \phi \over \sin^3
\phi }\pp \eta \pb \eta &=& 0 ,\nonumber \\
\pb (\cot^2 \phi ~\pp \eta ) + \pp (\cot^2 \phi ~ \pb \eta ) &=& 0.
\label{csg1}
\een
This reduces to the well-known sine-Gordon equation when $\eta =0$.

\section{The Complex sine-Gordon Model}

Since its first introduction by Lund and Regge  as explained in the
previous section, and also independently by Pohlmeyer \cite{pohlmeyer} in the context of the
reduced nonlinear sigma model\footnote{Eq.~(\ref{vortex}) can
be shown to be dual to the nonlinear sigma model \cite{nappi}, which
itself is also integrable \cite{zakharov,curtright,uhlenbeck}. The
vortex motion can be studied along this line. However, one has to impose
the constraint (\ref{constr}) later on, which is quite a nontrivial
task. In fact, this is why the gauged WZNW model is an appropriate framework where the constraints are
taken care of through the $U(1)$-gauging }, the complex sine-Gordon model has been studied extensively.
There exists a Lagrangian of the complex sine-Gordon equation
(\ref{csg}) in terms of $\phi$ and $\eta$ which however is singular at
specific values of $\phi$. This singularity problem has been resolved
beautifully by identifying the complex sine-Gordon model with the
integrably deformed $SU(2)/U(1)$-coset conformal model \cite{bakas2}.
The relevant action in general is given by a $G/H$-gauged
WZNW action plus a potential
term which accounts for an integrable deformation as follows \cite{Park};
\be
I = I_{WZNW}(g, A,\bar{A} )  - I_{P}(g, T, \Tb)
\label{WZWaction}
 \ee
where $I_{WZW}(g,A,\bar{A})$ is the usual gauged WZNW action for the $G/H$-coset conformal field
theory \cite{coset} with a map $g  : M \rightarrow G $ of a Lie group
$G$ defined on two-dimensional Minkowski space $M$ and gauge connections $A$ and $\bar{A}$
which gauge the anormaly free subgroup $H \in G$.  The deformation potential term is given in
terms of $T $ and $\Tb $,
\be
I_{P}(g, T, \Tb ) =  {\b \over 2\pi }\int \mbox{Tr}(gT\gi \Tb  ),
\ee
where $\b$ is a mass parameter and $T$ and $\Tb$ belong belong to the center of the
subalgebra $\bf{h} \subset \bf{g}$, i.e. $[T, \bf{h}] =[ \Tb, \bf{h}]=0$.
The gauged coset action $I_{WZNW}(g, A,\bar{A} )$ is characterized by the classical equation of
motion,
\be
\d _{g}(I_{WZNW}+ I_{P}) = {1 \over 2\pi }\int \mbox{Tr } (- [ ~ \pb +\bar{A} , ~
\pp + \gi \pp g  +\gi A g] +
\b [ ~ T , ~ \gi \Tb g ~ ]~ ) g^{-1}\d g = 0 ,
\label{zeroeqn}
\ee
and the constraint equation resulting from the variation of $I$ with respect to $A, \Ab $
\ben
\d _{A}I &=& {1 \over 2\pi }\int \mbox{Tr}( ( \ - \pb g
\gi + g\Ab \gi - \Ab \  )\d A )
= 0 , \nonumber \\
\d _{\Ab }I &=& {1 \over 2\pi }\int \mbox{Tr}( ( \  \gi
\pp g  +\gi A g - A \ )\d\Ab ) = 0 \ .
\label{constraint2}
\een
Or,
\be
( \ - \pb g \gi + g\Ab \gi - \Ab \  )_{\bf h} = 0 , ~ ~ ~
( \  \gi \pp g  +\gi A g - A )_{\bf h} = 0,
\label{const3}
\ee
where the subscript ${\bf h}$ specifies the projection to the subalgebra ${\bf h}$. It can be
readily checked that these constraint equations, when combined with Eq. (\ref{zeroeqn}), imply
the flatness of the connection $A$ and $ \Ab $, i.e.
\be
F_{z \zb } = [ \ \pp + A \ , \ \pb + \Ab \ ] = 0 \ .
\label{flat}
\ee
This flatness condition together with the vector gauge invariance of the action $I$ allow us to
set $A=\bar{A}=0$ so that the constraint equation simplies to
\be
(\gi \pp g  )_{\bf h} = 0, ~( \pb g \gi)_{\bf h} = 0 .
\label{gaugecon}
\ee
Using the identity,
\be
\pp ( \gi \Tb g ) + [ ~ \gi \pp g  \ ,  \  \gi \Tb g ~ ] = 0 ,
\ee
the equation of motion can be written as a zero curvature condition
\be
[~ \pp + \gi \pp g +  \b \l T \ , \ \pb  +{1 \over \l } \gi
\Tb g ~ ] = 0  ~ ,
\label{zero}
\ee
where  $\l $ is an arbitrary spectral parameter. In turn, this zero curvature condition can be
understood as the integrability condition of the linear system
\be
(\pp + U )\Phi = 0 \ ,\ \
(\pb + V )\Phi = 0,
\label{linear}
\ee
where $U \equiv U_0 + \l U_1 =\gi \pp g   +  \b \l T$ and
$V \equiv  {1 \over \l} V_1 ={1 \over \l } \gi \Tb g.$

Now we restrict to the complex sine-Gordon case where the coset $G/H$ is $SU(2)/U(1)$.
For $g \subset SU(2)$ element,  we parametrize $g$ by
\be
g = \pmatrix{ u & i\sqrt{1-u\ub }e^{ i \q }  \cr
i\sqrt{1-u\ub } e^{ -i \q} & \ub },
\ee
and take Pauli matrices $\s_i$ as generators of $\bf{g}$. Also, we assume that
$T=-\Tb = i \s_ 3$. Then, the gauge constraint (\ref{gaugecon}) can be written as
\ben
0 &=& {\ub \pp u - u\pp \ub \over 4(1-u\ub )} - {i \over 2}\pp \q,
\nonumber  \\
0 &=& {u\pb \ub - \ub \pb u \over 4(1-u\ub )} - {i \over 2}\pb \q ,
\label{newcon}
\een
which may be used to bring Eq. (\ref{zero}) into a more conventional form of the
complex sine-Gordon equation,
\ben
\pp\pb u + {\ub \pp u \pb u \over 1-u\ub } + 4\b u(1-u\ub ) &=& 0, \nonumber \\
\pp\pb \ub + {u \pp \ub \pb \ub \over 1-u\ub } + 4\b \ub (1-u\ub ) &=& 0 .
\label{csg}
\een
By writing $u=\cos{\phi}\exp(i\eta)$, one can readily see that this equation indeed agrees
with Eq. (\ref{csg1}). The zero curvature condition in Eq. (\ref{int}) also agrees with
Eq. (\ref{zero}) up to $U(1)$-vector gauge transformation. Since the complex sine-Gordon
equation and the vortex equation are both invariant under the vector gauge transformation,
we adhere to the gauge, $A=\bar{A}=0$ for simplicity for the rest of the paper.

Having formulated the complex sine-Gordon equation  as the
integrability condition of the linear equation, we may obtain the B\"{a}cklund transformation
of the complex sine-Gordon equation which generates a new solution from a given solution
as follows; if $(f, \F_{f})$ is a solution of the linear equation such that
\be
(\pp + f^{-1} \pp f +\b \l T)\F_{f} = 0 \ ,  ~~~
(\pb + {1 \over \l}f^{-1} \Tb f )\F_{f} = 0 ,
\label{newlin}
\ee
a new set of solution $(g, \F_{g})$ is given by
\be
\F_g = {\l \over \l-i \d / \sqrt{|\b|}} \left( 1-{i\d \over \l \sqrt{|\b|}} g^{-1} \s_{3} f \right)
\F_f ,
\label{bt1}
\ee
with an arbitrary parameter $\d$, provided that $g$ and $f$ satisfy the B\"{a}cklund transformation,
\ben
\gi \pp g - f^{-1}\pp f - \d \sqrt{|\b|} \ [\ \gi \s_{3} f \ ,
\ \s_{3} \ ] &=& 0  , \\
\d \pb (\gi \s_{3} f) +  \sqrt{|\b|}\  \gi \s_{3} g - \sqrt{|\b|}
\ f^{-1} \s_{3} f &=& 0  .
\label{bt2}
\een
Specifically, we will be concerned about the following four distinct types of solutions of the
complex sine-Gordon equation, which we dualize later to the vortex solutions.
First, we consider the trivial vacuum case. For $\b <0$ which we assume below without loss of generality,
the vacuum solution is given by
\ben
\mbox{\bf Case A:} ~~~~~~f&=& 1 , ~u=1, \nonumber \\
\Phi&=& \Phi_f =\pmatrix{ e^{-i \b \l z +i \zb / \l} & 0  \cr
0 & e^{i \b \l z -i \zb / \l }}.
\label{tri}
\een
Note that $\theta $ can be of any value so that the vacuum has an unbroken $U(1)$-symmetry.

A less trivial solution can be obtained if we make an assumption of the constant magnitude
of $u$, i.e. $|u|=\a $,
\ben
\mbox{\bf Case B:}~~~~~~
u&=& \a \exp \left(i (\a^2-1) (\g z -{\zb \over \g} ) \right), ~~ 0 \le \a <1 , \nonumber \\
\q&=& -\a^2 (\g z +{\zb \over \g})
\label{smoke2},
\een
where $\a$ and $\g$ are arbitrary constants. The linear function $\Phi_{f}$ will be found
later when we consider the corresponding vortex solution.

In order to obtain the one soliton solution, we need
to apply the B\"{a}cklund transformation to the vacuum solution (Case A) and impose the gauge constraint.
For $\b < 0$ and $f=1$,  the components of Eq.(\ref{bt2}) reduce to
\ben
\pp u + 2\d \sqrt{-\b }(1-u\ub ) &=& 0, \nonumber \\
\pb u - { 2  \sqrt{-\b } \over  \d }(1-u \ub ) &=& 0,
\een
and their complex conjugates. These equations
may be readily integrated to yield the 1-soliton solution,
\ben
\mbox{\bf Case C:}~~~~~~
u &=& -\cos {\a } \tanh \left(2 \sqrt{-\b }
\cos {\a } (\d z- {\zb \over \d}) \right)  - i \sin {\a },
\nonumber \\
\q &=& -2 \sqrt{-\b}  \sin {\a }  (\d z+ {\zb \over \d}),
\label{1sol}
\een
where
$\cos{\a } $ and $ \sin{\a }$ ($ -\pi \le \a < \pi $)
are constants of integration and
\be
z=-{1 \over 8 \l}(\t +\s), ~ ~ \zb =-{\b \l \over 8} (\t - \s) .
\ee
Note that this soliton solution reduces to the famous sine-Gordon kink solution when $\a =0$.
When $\a \ne 0$, the above soliton becomes nontopological and carries an extra conserved
$U(1)$-charge \cite{Hol,Shin}. In particular, if $\a = \pi/2$, it simply reduces to the vacuum
solution. Thus, the 1-soliton in Eq. (\ref{1sol}) interpolates between the topological kink and
the vacuum solution.

Two-soliton solution and the breather solution can be obtained by the
following nonlinear superposition rule; let $g_1$ and $g_2$ be a pair of
one soliton solutions obtained through the B\"{a}cklund transformation applied to the
trivial solution $f=1$, with parameters of transformation $\d_1$ and $\d_2$ respectively.
Then, by taking a successive application of the B\"{a}cklund transformation to each one-solitons
but with parameters reversed, i.e. $\d_2$ and $\d_1$, and requiring the commutability of each processes
we obtain a nonlinearly superposed solution $g_3$ such that
\be
g_3 = \s_{3} (\d_2 g_1 -\d_1 g_2) g_0^{-1} \s_{3} (\d_1 g_1^{-1}
-\d_2 g_2^{-1} )^{-1}.
\ee
This superposed solution in general describes the scattering of two solitons. For example, it describes
a soliton-soliton scattering if $\d_{1}=-1/\d_{2}$, and a soliton-antisoliton scattering if $\d_{1}= 1/\d_{2}$
for real $\d_{1}$ and $\d_{2}$. If we assume $\d_{1}$ and $\d_{2}$ to be unit complex numbers such that
$\d_1=1/\d_{2} =\d$ with $|\d |=1$, we obtain a breather solution of the complex sine-Gordon
equation \cite{Shin}. Here, for simplicity, we consider only the breather of the sine-Gordon equation which
arises as a special case of breathers of the complex sine-Gordon equation by considering the superposition of
topological solitons. In our notation, if we denote
\be
g_i = \exp(i \s_2 \f_i); \ \ i=1,2,3,
\ee
where $\f_1 = 2 \tan^{-1} \exp(2 \sqrt{-\b} (\d z - \zb / \d  ))$,
$\f_2 = 2 \tan^{-1} \exp(2 \sqrt{-\b} (- \d \zb + z/\d  ))$, the breather solution is given by
\ben
\mbox{\bf Case D:}~~~~~~ u &=& \cos{\phi_{3}}, ~ \theta = {\pi \over 2}, \nonumber \\
\f_3 &=& \tan^{-1} {(\d^2 - 1 / \d^2 ) \sin (\f_1-\f_2)
\over(\d^2 + 1 / \d^2 ) \cos (\f_1-\f_2) -2}.
\label{breather}
\een
In the next section, vortex configuration corresponding to these solutions will be constructed.

\section{Dualization and vortex solutions}
In Sec. 2, the vortex model of Lund and Regge has been dualized to the the complex sine-Gordon model by writing
the complex sine-Gordon variables in terms of vortex coordinate variables and showing that the vortex equation
implies the complex sine-Gordon equation. The reverse direction, i.e. writing the vortex variables in terms
of the complex sine-Gordon variables, was not known in the original work by Lund and Regge, and has been completed
only recently \cite{Park2}. This can be stated easily in terms of the associated linear equations in Eq. (\ref{linear})
as follows; let $\Phi (z, \bar{z}, \l ) $ be a solution of the linear equation of the complex sine-Gordon model.
Then, the matrix $F$ defined by
\be
F  \equiv \Phi ^{-1} \l {\pp \over \pp \l} \Phi = \Phi ^{-1} {\pp \over \pp t} \Phi; ~ (t \equiv \ln \l )
\label{fdef}
\ee
results in the vortex coordinate through
\be
F \equiv \a \sum_{i=1} ^3 i X_i \s_i ,
\label{vorcoor}
\ee
up to some normalization constant $\a$. In order to prove that these  $X_{i}$ indeed satisfy the vortex equation as
well as the constraint equation, we first compute that
\be
\pp F = -\Phi ^{-1} \pp \Phi \Phi ^{-1} {\pp \over \pp t} \Phi +
\Phi ^{-1} {\pp \over \pp t} (\pp \Phi) = -\Phi ^{-1} {\pp U \over \pp t} \Phi
= - \l \Phi ^{-1} U_1 \Phi.
\ee
In a similar way, we find that $\pb F={1 \over \l} \Phi ^{-1} V_1 \Phi$.  This shows that the
constraint equation is automatically satisfied,
\be
\tr (\pp F)^2 = 2 \a^2 (\pp \X)^2 = -2 \b^2 \l^2 ,\ \
\tr (\pb F)^2 = 2 \a^2 (\pb \X)^2 = -{2 \over \l^2} ,
\ee
up to some rescaling to be determined later.
Since
\ben
\pb \pp F &=& \pb (-\Phi^{-1} {\pp U \over \pp t} \Phi)
=\Phi ^{-1} ([{\pp U \over \pp t}, V] - \pb {\pp U \over \pp t} ) \Phi ,\nonumber \\
\ \ \pp \pb F &=& \pp (-\Phi^{-1} {\pp V \over \pp t} \Phi)
=\Phi ^{-1} ([{\pp V \over \pp t}, U] - \pp {\pp V \over \pp t} ) \Phi ,
\label{comp}
\een
we note that the compatibility condition: $\pp \pb F = \pb \pp F$ requires that
\be
\pp V - \pb U +[U, V] = [\pp + U\ , \pb + V] = 0 .
\ee
This is just the zero curvature equation (\ref{zero}) which holds when the complex
sine-Gordon equation is satisfied. Thus, one can find such a matrix $F$ for any given solution
of the complex sine-Gordon equation.
In particular, the first order term in $\l$ in the zero curvature expression yields
$-\pb U_1 = 0$. Using this fact in Eq. (\ref{comp}), it is easy to verify that
\be
\pb \pp F = \Phi ^{-1} [ U_1,\  V_1] \Phi = -[\pp F \ , \pb F].
\label{com}
\ee
In terms of the component as in Eq. (\ref{vorcoor}), this reduces precisely to the vortex equation
\begin{equation}
   \pp^{2}_{\tau} X_i -\pp^{2}_{\s } X_i  -  \b \e_{ijk} \pp_{\tau } X_j \pp_{\s } X_k = 0 ,
\end{equation}
when we rescale world-sheet coordinates and fix the constant $\a$ in the following way;
\be
z \rightarrow -{1 \over 4 \l} z, ~~ \zb \rightarrow -{\b \l \over 4} \zb, ~~
\a \rightarrow {i \b \over 4}.
\label{scale}
\ee
More explicitly, $X_i$ are given by
\be
X_i = {4 i \over \b} F_i\Big( z=-{1 \over 8 \l}(\t +\s), \zb =-{\b \l \over 8} (\t - \s) \Big) .
\label{realco}
\ee
Thus, the vortex coordinate $X_i$ can be obtained explicitly by solving the associate linear
equation of the complex sine-Gordon equation, thereby completing the dualization procedure.

Besides the duality between the equations,  the dualization of the two models in terms of an
integrably deformed gauged WZNW action provides dualities between the global properties, e.g.
symmetries of the model.
Since the vortex model in our case is relativistic, it possesses world-sheet Lorentz
invariance, and also a discrete symmetry under the interchange $\s \leftrightarrow \t$.
This obviously contrasts with the usual case of vortex equation in fluid dynamics in Eq. (\ref{Bet})
where  the discrete symmetry is absent. In the context of the deformed WZNW action, the discrete symmetry
corresponds to the parity symmetry under the action $g \rightarrow \h g,\ \ \zb \rightarrow - \zb$
where $\h$ is a generator which anti-commutes with $\s_{3}$.

Another important symmetry comes from the symmetry of the linear equation
in Eq. (\ref{linear}) under the action $\Phi(\l, z, \zb) \rightarrow
\Phi(\l, z, \zb) \tilde{\Phi}(\l)$ for any arbitrary matrix function $\tilde{\Phi}(\l)$. This induces
the rotational and the translational transformation of vortex coordinates such that
\be
F \rightarrow \tilde{\Phi}(\l)^{-1} F \tilde{\Phi}(\l) + \l \tilde{\Phi}(\l)^{-1}
{d \tilde{\Phi}(\l) \over d \l}.
\label{symme}
\ee
The local $U(1)$-gauge symmetry of the deformed gauged WZNW action does not contribute to the
vortex model. Under the $U(1)$-gauge transformation,  $F$ as defined in Eq. (\ref{fdef}) is gauge invariant.
Thus, in the following, we fix the gauge via the gauge constraint: $(g^{-1} \pp g)_{\bf{h}}
=(\pb g g^{-1})_{\bf{h}}=0$ and compute vortex solutions by dualizing solutions (Cases A-D in
Sec. 3) of the complex sine-Gordon equation.
\subsection{Case A: Straight Line}
Calculating $F$ using $\Phi_f$ in Eq. (\ref{tri}), we have
$F=i \s_3 (- \b \l z - \zb /\l)$. Thus, Eq. (\ref{realco}) gives rise to
$X_1 = X_2 =0,\ \ X_3= -\s$, which describes a straight line. This straight line can be mapped
to any other line through the symmetry action in Eq. (\ref{symme}).

\subsection{Case B: Kelvin wave and smoke ring}
The smoke ring solution in fluid dynamics has the form,
\be
\X = \left( A(\t) \cos(c \s), A(\t) \sin(c \s), B(\t) \right).
\ee
Note that when $A(\t )$ is a constant, this arises as a special case of a Kelvin wave in fluid dynamics
of the form,
\begin{equation}
\X = \left(A\cos(k \sigma -\omega \tau), A\sin(k\sigma -\omega \tau),
B\sigma\right),
\end{equation}
if we interchange $\s $ and $\t$ using the discrete symmetry and fix $k=0$.

Now, we show that the vortex solution corresponding to Case B has the form of a Kelvin wave.
First, in order to solve for $\Phi$ with the solution $u$ and $\q$ in Eq. (\ref{smoke2}),
we define a new variable $\hat{\Phi} \equiv \Phi_0 ^{-1} \Phi$ where
\be
\Phi_0 =\pmatrix{ exp(-{i \over 2} \D) & 0  \cr
0 & exp({i \over 2} \D)} ,\ \ \D = (2\a^2 -1) \g z+{\zb \over \g}.
\ee
Then the linear equation (\ref{newlin}) changes into
\ben
\left(\pp + (g^{-1} \pp g)_{\mbox{new}} -{\l \over 4} +({i \over 2} \g -i \g \a^2) \s_3 \right)
\hat{\Phi} &=& 0, \nonumber \\
\left(\pb + {1 \over \l}(g^{-1} \Tb g)_{new} - {i \over 2 \g} \s_3 \right)\hat{\Phi} &=& 0,
\label{newlinear}
\een
where
\ben
(g^{-1} \pp g)_{\mbox{new}}&=&  \pmatrix{ 0 &  \g \a \sqrt{1-\a^2} \cr
- \g \a \sqrt{1-\a^2} & 0 }, \nonumber \\
(g^{-1} \Tb g)_{\mbox{new}} &=& \pmatrix{
-i (2 \a^2-1) & 2  \a \sqrt{1-\a^2} \cr
- 2  \a \sqrt{1-\a^2} & i (2 \a^2-1) }.
\een
One can readily integrate Eq. (\ref{newlinear}) with the result,
\be
\hat{\Phi} = \cos Y + i \hat{\s} \sin Y,
\ee
where
\be
Y=-\sqrt{A^2+B^2} (z + {2 \over \l \g} \zb ),\ \
\hat{\s} = (A \s_3 + B \s_2 ) /  \sqrt{A^2+B^2},
\ee
and
\be
A = - {\l \over 4} + {\g (1 - 2 \a^2) \over 2},\ \
B = \sqrt{1 - \a^2} \g \a.
\ee
Then, the matrix $F$, followed by a rotation as explained in Eq. (\ref{symme}) such that
$F= M {\Phi}^{-1}\l d {\Phi} / d \l$ and
\be
M = \pmatrix{ 1 & 0 & 0 \cr
0 & A/\sqrt{A^2+B^2} & -B/\sqrt{A^2+B^2} \cr
0 & B/\sqrt{A^2+B^2} & A/\sqrt{A^2+B^2} },
\ee
now becomes
\ben
F_1 &=&{ i \lambda B \cos 2Y \over 8 (A^2+B^2)}, \nonumber \\
F_2 &=& { i \lambda B \sin 2Y \over 8 (A^2+B^2)}, \nonumber \\
F_3 &=& {i \over 4 \g \l} {8(A^2+B^2) \zb +A \l (\g \l z+2 \zb)  \over
\sqrt{A^2+ B^2}}.
\een
Finally, we obtain the solution in terms of the vortex coordinates $X_i$ using Eq. (\ref{realco}),
\ben
X_1 &=& -{  \lambda B \cos 2Y \over 2 \beta (A^2+B^2)}, \nonumber \\
X_2 &=&- {  \lambda B \sin 2Y \over 2 \beta (A^2+B^2)}, \nonumber \\
X_3 &=& \Big( -{1\over \g} +{A\g -2\b\l A \over 8\b\g (A^2+B^2)}\Big) \s
+ \Big( {1\over \g} +{A\g +2\b\l A \over 8\b\g (A^2+B^2)}\Big) \t
, \nonumber \\
Y &=& { \sqrt{A^2+B^2} \over 8 \l \g }\Big( (\g - 2\b \l )\s + (\g + 2\b \l )\t \Big).
\een
This agrees with a Kelvin wave after an appropriate world-sheet coordinate
change through Lorentz transformation.
In particular, if we take $ \gamma =\lambda /2$ and $ \beta = -1/4$ as a special case, we have
\ben
X_1 &=&{ 4\over \alpha} \sqrt{1-\alpha^2} \cos{\alpha \sigma \over 4} , \nonumber \\
X_2 &=&  {4\over \alpha }\sqrt{1-\alpha^2}\sin{ \alpha \sigma\over 4}, \nonumber \\
X_3 &=&  \a \t ,
\label{smoke}
\een
where $0 \le \a <1$. This indeed has the form of a smoke ring.
\begin{figure}
\leftline{\epsfxsize 1.8 truein \epsfbox {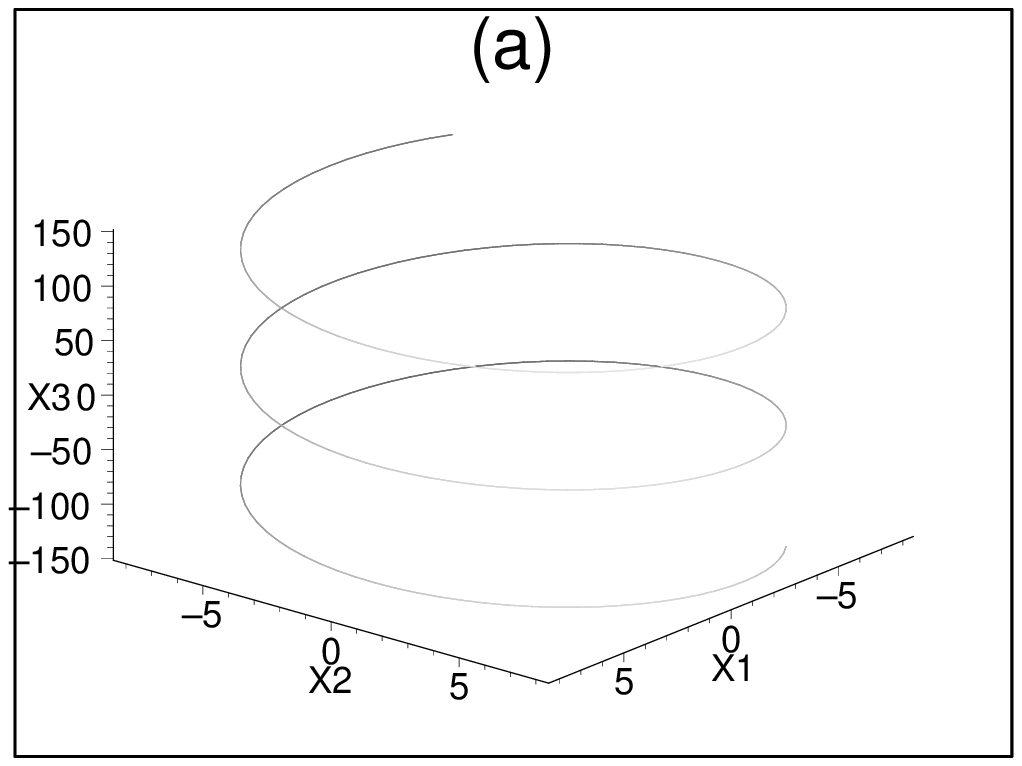}}
\vglue -1.365 in
\centerline{\epsfxsize 1.8 truein \epsfbox {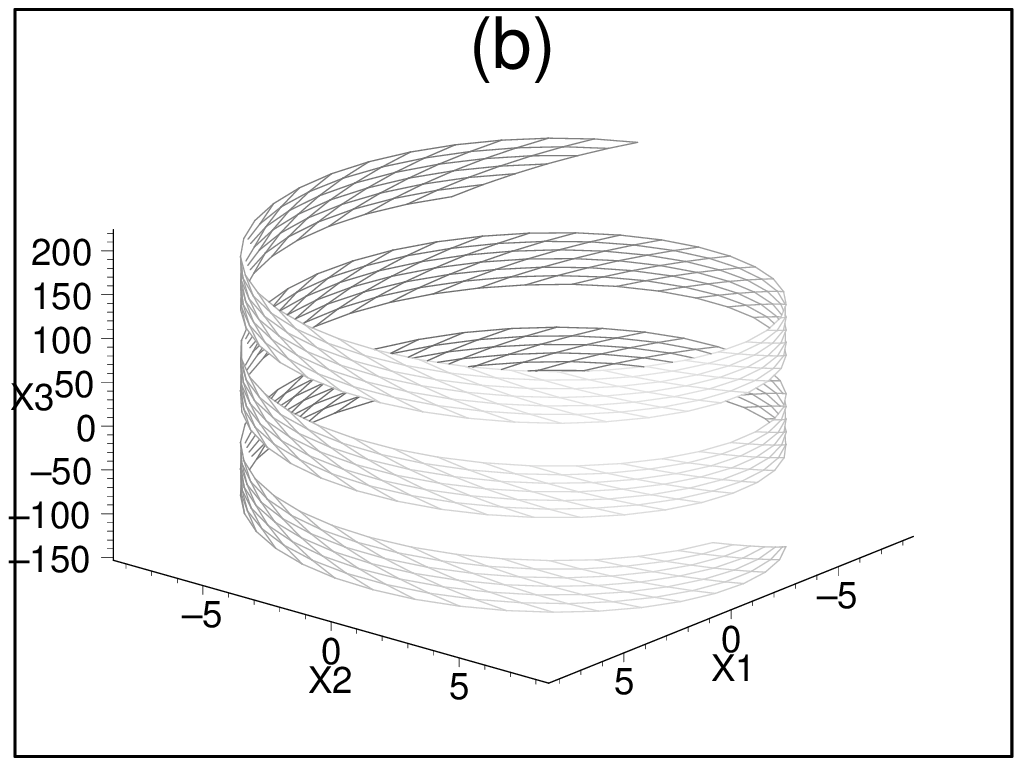}}
\vglue -1.365 in
\rightline{\epsfxsize 1.8 truein \epsfbox {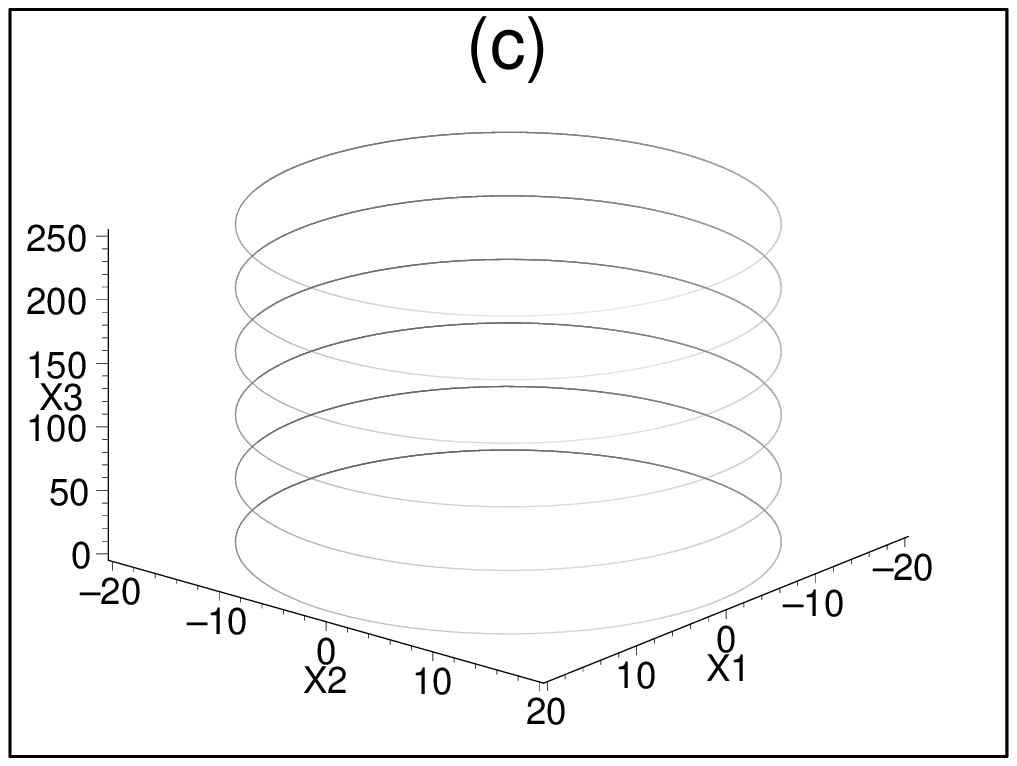}}
\vglue 0.2in
\caption{(a) A Kelvin wave at $\tau =0$, (b) and its motion during $-7 < \tau <7$.  (c) Smoke
rings at $\t = 0, 10, ...,40, 50$. }
\end{figure}
Figure 1 shows a Kelvin wave and a smoke ring configurations. Fig. 1-(a) is a Kelvin wave at $\t =0$ and
Fig. 1-(b) shows its motion during $-7 < \tau <7$ moving along the z-axis. Fig. 1-(c) shows smoke rings
at $\t = 0, 10, ...,40, 50$. It shows that the surface swept out by
a smoke ring is a cylinder. The straight line of Case A is thus a limiting case where the ring
shrinks to a point.

\subsection{Case C: Soliton}
One of the well-known vortex configurations is the solitary kink wave which propagates along
a straight vortex filament line. This was first found by Hasimoto \cite{hasimoto} using the
duality between the Da Rios-Betchov equation and the nonlinear Schr\"odinger equation.
This Hasimoto soliton, which was obtained by dualizing the NLS soliton, has the form,
\be
\X = A \rm{sech}(\a \s -\b \t) \left( \cos(\g \s -\d \t), \
\sin(\g \s -\d \t), \ 0\right) + \left( 0,\ 0,\ B\s -A \tanh(\a \s -\b \t) \right).
\ee
In the present case of the Lund-Regge model, a similar configuration can be found also by
dualizing the soliton of the complex sine-Gordon equation as given in Case C.
In order to do so, we first compute $\Phi$ by using the B\"{a}cklund transformation in
Eq. (\ref{bt1}) applied to the vacuum ($f=1$) so that
\be
\Phi= {\l \over \l-i \d / \sqrt{-\b }} \Phi_1 \Phi_f,
\ee
where $\Phi_f$ is given in Eq. (\ref{tri}) and
\be
\Phi_1=1+ {\d \over \sqrt{-\b} \l} \pmatrix{ -iu^* & \sqrt{1-u u^*} e^{i \q}  \cr
-\sqrt{1-u u^*} e^{-i \q} & i u },
\ee
with $u$ and $\q$ as in Eq. (\ref{1sol}).
A straightforward calculation using Eq. (\ref{fdef}) shows that
\ben
F_1 &=& i R \sqrt{1-u u^*}\  \sin(2 \b \l z - {2 \over \l} \zb + \q), \nonumber \\
F_2 &=& i R \sqrt{1-u u^*}\  \cos(2 \b \l z - {2 \over \l} \zb + \q), \nonumber \\
F_3 &=& {i R \over 2} (u + u^* ) - i \l \b z - {i \over \l} \zb, \nonumber \\
R &=& { \sqrt{-\b} \l \d \over  -\b \l^2 + \d^2 + 2 \sqrt{-\b} \l \d \sin{\a }}.
\een
Using Eq. (\ref{scale}), we obtain the soliton solution,
\ben
X_1 &=& {4R \over \b} \cos{\a }\  \rm{sech} \S \ \sin (-{\b \over 2} \s + \q), \nonumber \\
X_2 &=& {4R \over \b} \cos{\a}\  \rm{sech} \S \ \cos (-{\b \over 2} \s + \q), \nonumber \\
X_3 &=& -{4R \over \b} \cos{\a } \ \tanh \S - \t,
\label{hasi}
\een
where
\ben
\S &=&  {1\over 4} \sqrt{-\b} \cos{\a } \Big( ({\d \over \l }-{\b \l \over \d})\t+( {\d \over \l }
+{\b \l \over \d})\s \Big), \nonumber \\
\theta&=& {1\over 4} \sqrt{-\b} \sin{\a } \Big( ({\d \over \l }+{\b \l \over \d})\t+({\d \over \l }
-{\b \l \over \d})\s \Big),
\een
which has the same form with the Hasimoto soliton in fluid dynamics.

\begin{figure}
\leftline{\epsfxsize 1.8 truein \epsfbox {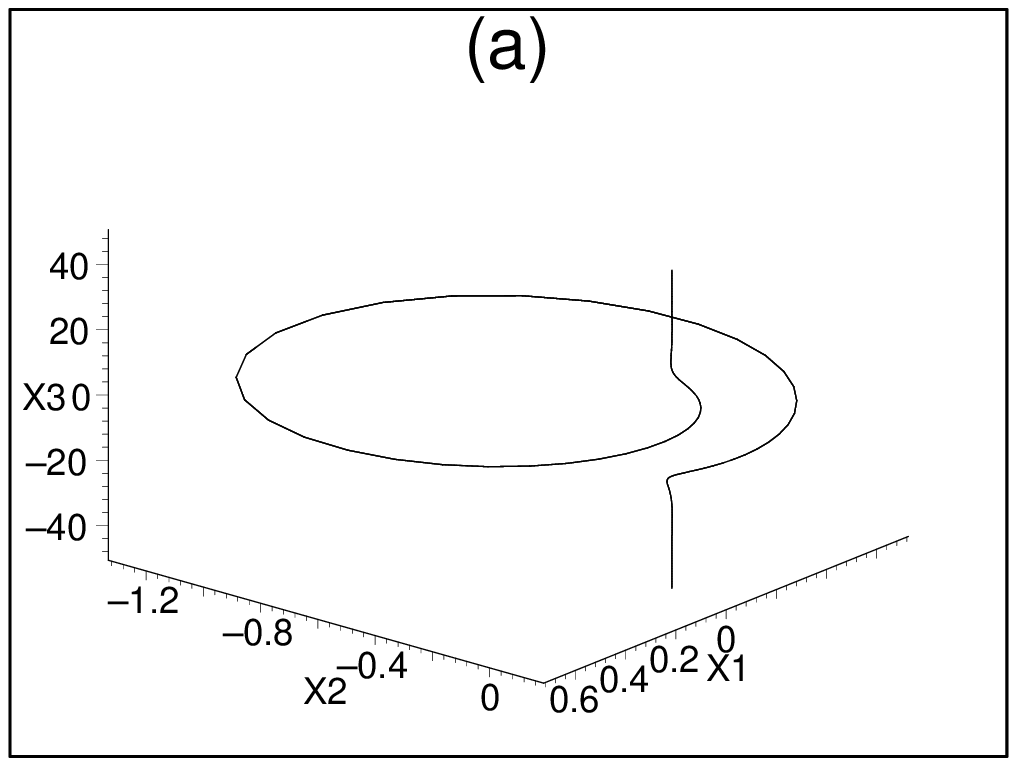}}
\vglue -1.365 in
\centerline{\epsfxsize 1.8 truein \epsfbox {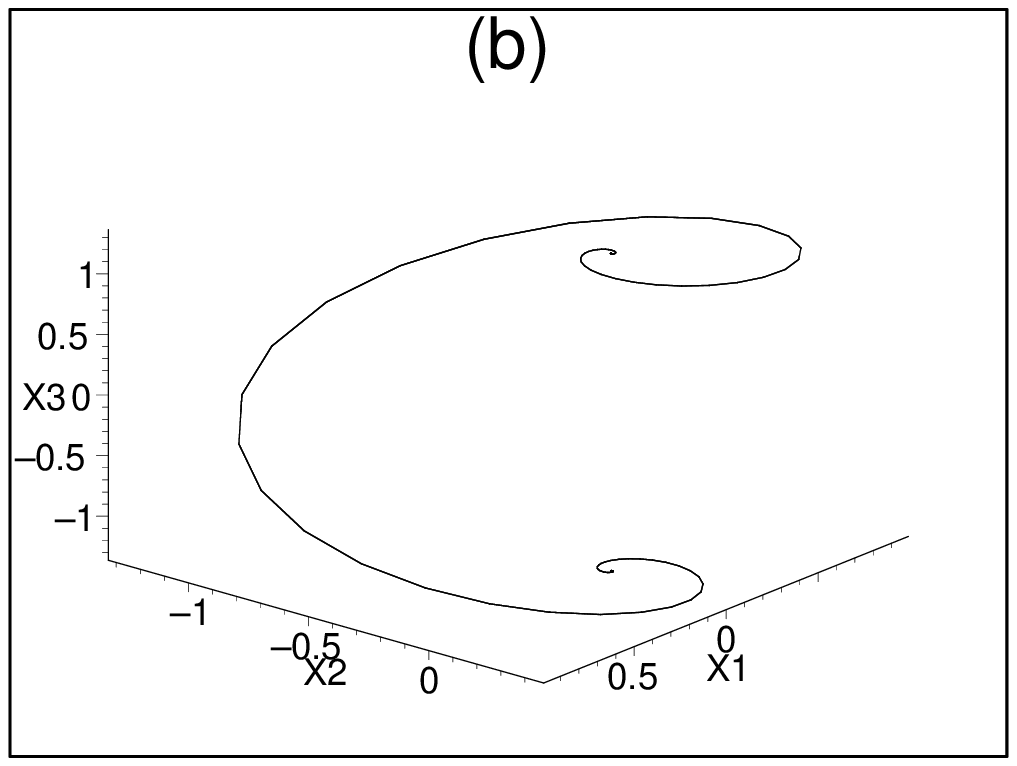}}
\vglue -1.365 in
\rightline{\epsfxsize 1.8 truein \epsfbox {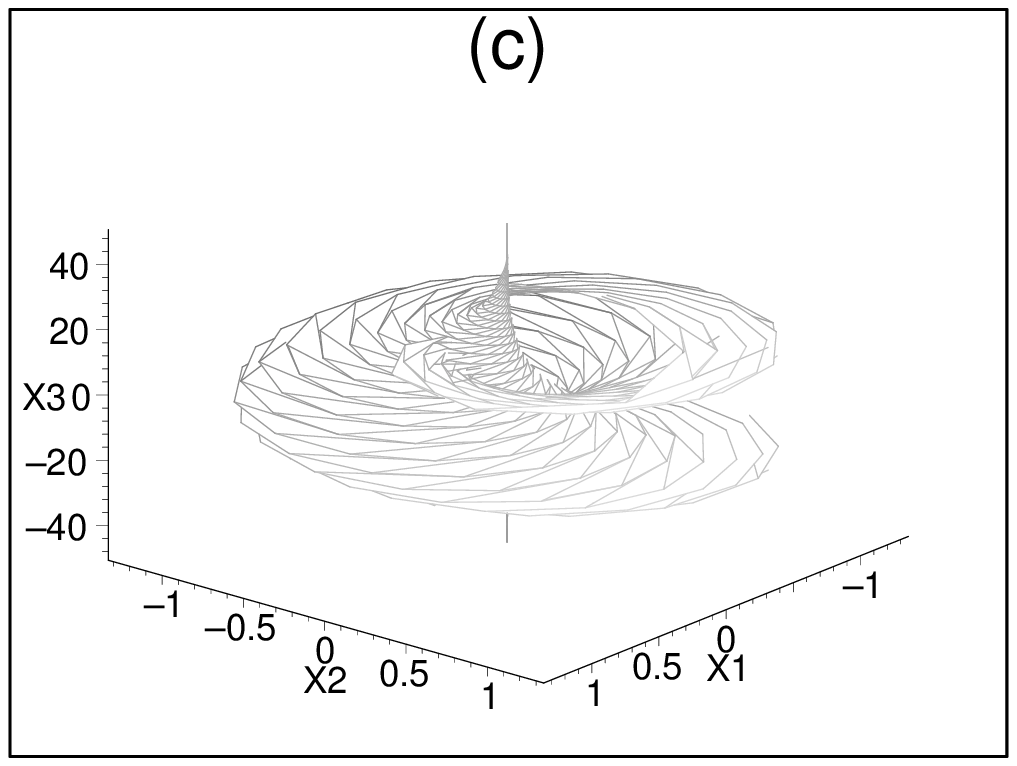}}
\vglue 0.2in
\caption{(a) A soliton vortex configuration at $\sigma =0$ and (b) an open string at $\tau =0$.
(c) An embedded surface for $-20 < \s <20$ and $-50 <\tau <50$.}
\end{figure}
Figure 2 shows soliton vortex configurations with parameters
$\a=0.8$, $\b=-0.3$, $\d=1.5$ and $\l=0.5$. Fig. 2-(a) is a Hasimoto-type vortex configuration where $\s =0$
and the curve is parametermized by $\tau$. Fig. 2-(b) is a open string configuration where $\tau =0$ and the
curve is parametrized by $\s$, and Fig. 2-(c) is a surface swept out by a moving soliton vortex.
We emphasize that due to the discrete symmetry of the vortex equation under the exchange $\s \leftrightarrow
\t$, the solution in Eq. (\ref{hasi}) with $\s$ and $\t$ exchanged is also a solution, thus roles of $\s$ and
$\t$ in the Figure 2 can also be exchanged.
\subsection{Breather vortex}
The breather solution of the sine-Gordon equation is presumably the most well-known
configuration in soliton theories as a stable localized solution except solitons.
Our dualization procedure suggests that we can also have a counterpart of a breather in vortex
dynamics.
As far as we know, such a configuration is not known in vortex dynamics literatures.
In order to construct the solution explicitly, we use the BT in Eq. (\ref{bt1}) to find $\F$ of the
linear equation (\ref{linear}),
\ben
\F &=& (1+\D_3 g_3^{-1} \Tb g_1)(1+\D_1 g_1^{-1} \Tb ) \F_f \nonumber \\
    &=&  \left( 1-i \s_3 \D_3 e^{i \s_2 (\f_1+\f_3)} \right)
 \left( 1-i \s_3 \D_3 e^{i \s_2 \f_1} \right) \F_f,
\een
where $\D_1 = {\d \over \l \sqrt{-\b}}$, $\D_3 = {1 \over \d \l \sqrt{-\b}}$
and $\f_1, \f_3$ and $\F_f$ are given by Eq. (\ref{breather}) and (\ref{tri}).
Using the definition $F \equiv \Phi^{-1} {\pp \over \pp t} \Phi$, we obtain
\ben
F_1 &=& {i \D_3 \over (1+\D_1^2)(1+\D_3^2)} \left( \D_1^2 \sin(\f_1-\f_3) \cos 2Y
+\sin(\f_1+\f_3) \cos 2Y - 2 \D_1 \sin \f_3 \sin 2Y \right) \nonumber \\
&+& {i \D_1 \over 1+\D_1^2} \sin \f_1 \cos 2Y, \nonumber \\
F_2 &=& {i \D_3 \over (1+\D_1^2)(1+\D_3^2)} \left( -\D_1^2 \sin(\f_1-\f_3) \sin 2Y
-\sin(\f_1+\f_3) \sin 2Y - 2 \D_1 \sin \f_3 \cos 2Y \right) \nonumber \\
&-& {i \D_1 \over 1+\D_1^2} \sin \f_1 \sin 2Y, \nonumber \\
F_3 &=& {i \D_3 \over (1+\D_1^2)(1+\D_3^2)} \left( \D_1^2 \cos(\f_1-\f_3)
+\cos(\f_1+\f_3) \right) + {i \D_1 \over 1+\D_1^2} \cos \f_1 -i\l \b z - {i\zb \over \l },
\een
where $Y=\b \l z -{\zb \over \l}$. This gives a ``breather vortex solution" after using the
relation in Eq. (\ref{realco}).  Figure 3 describes breather vortex configurations
and the embedded surface. Fig. 3-(a) shows an open-string type breather vortex with $\t =0$.
Fig. 3-(b) a breather vortex configuration with $\s =0$, and Fig. 3-(c) shows an embedded
surface. Note that Fig. 3-(a) shows a cusp at $\s =0$. However, this does not cause any physical
singularity which may be seen by checking the derivatives of vortex coordinates
with respect to $\s$ at $\s=0$. Explicit calculation shows that they behave regularly at $\s =0$.
In fact, this cusp-like behavior arises since we have restricted to the breather of the sine-Gordon equation
which arises as a special case of breather solutions of the complex sine-Gordon equation.
Instead of proving this in terms of detailed solutions, here we only content that breather solutions obtained
from solitons with $\a \ne 0$ do not possess this cusp-like behavior.
Besides, we may expect that this new type vortex
configuration is stable since it arises from dualizing a breather solution of the complex sine-Gordon
equation which is stable. However, this has to yet to be seen.
\begin{figure}
\leftline{\epsfxsize 1.8 truein \epsfbox {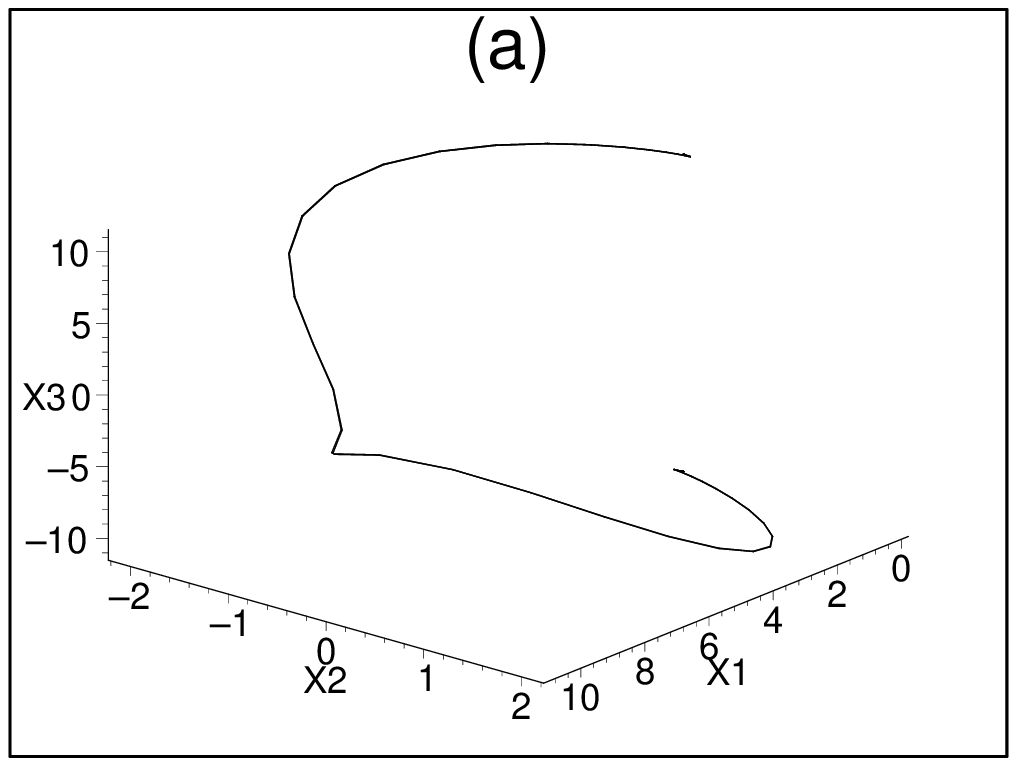}}
\vglue -1.365 in
\centerline{\epsfxsize 1.8 truein \epsfbox {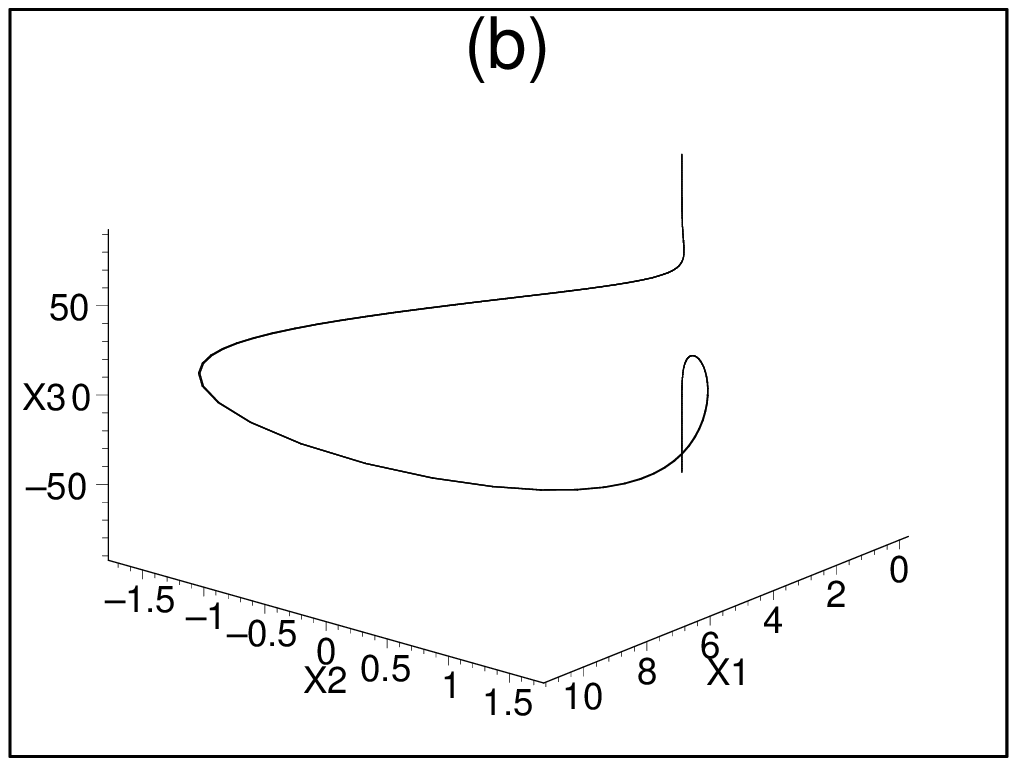}}
\vglue -1.365 in
\rightline{\epsfxsize 1.8 truein \epsfbox {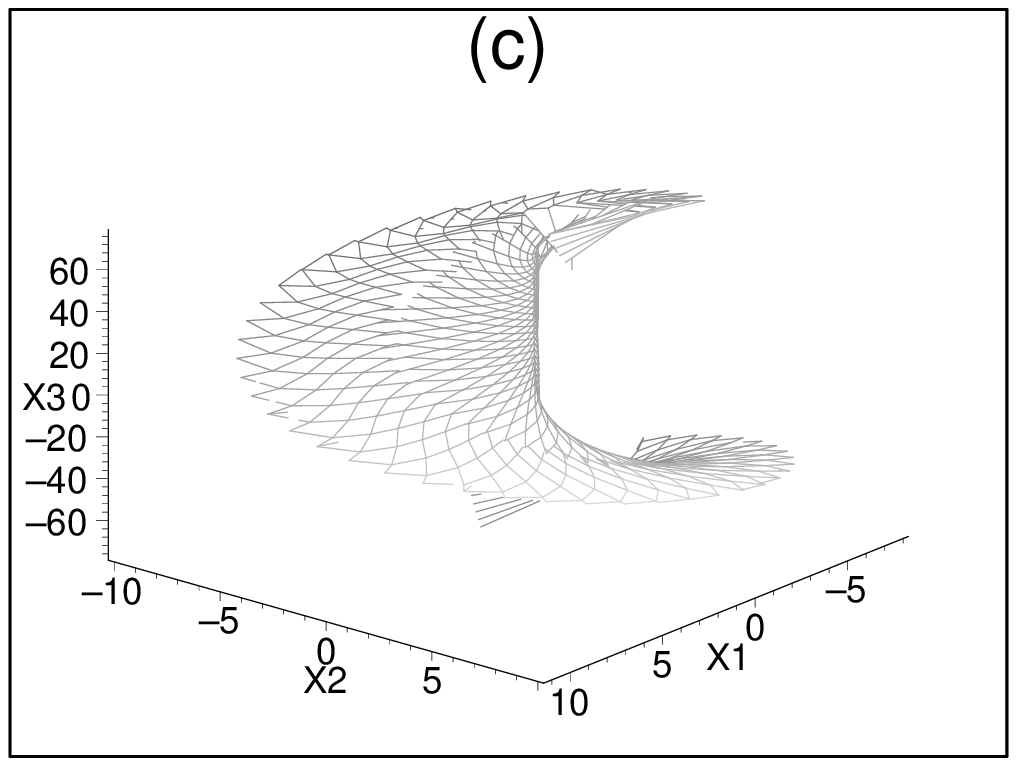}}
\vglue 0.2in
\caption{Breather vortex configurations with parameters $\l=0.5$, $\b=-0.09$, $\d=e^{0.5 i}$, (a) an open-string
configuration with $\t =0$ and (b) a breather vortex with $\s =0$. (c) Embedded surface for $ -100 <\tau  <100$
and $-60 < \s < 60$.}
\vglue 0.5in
\end{figure}

\section{Discussion}
In this paper, we have found various vortex configurations of the Lund-Regge model by making use
of the duality between the Lund-Regge equation and the complex sine-Gordon equation.
We have shown that these vortices have analogous counterparts in the vortex system of fluid
dynamics. In fact, these correspondence arise from the fact that the usual fluid vortex equation
is a ``nonrelativistic limit" of the Lund-Regge equation. This may be also seen in their
dualized versions, i.e. the nonlinear Schr\"{o}dinger equation which dualizes the vortex
equation in fluid dynamics also arises as a nonrelativistic limit of the sine-Gordon
equation, which itself is a special case of the complex sine-Gordon equation. The exact
correspondence of these equations will be considered elsewhere.

Another important problem would be the quantization of the system. In view of
the string formulation of the system by Lund and Regge, the quantization of the present case would be a stringy
generalization of the Landau levels of a charged particle moving on a uniform magnetic
field. The magnetic flux vortex in the Maxwell Higgs system with a uniform magnetic field, for
exampe, can be quantized consistently. It would be interesting to know whether the duality between the Lund-Regge
model and the complex sine-Gordon model persists at the quantum level.
\vskip 2cm
\noindent{\Large \bf Acknowledgements}
\vskip 1cm
The work of K.Lee was supported in part by the SRC program of the SNU-CTP, the Basic Science and
Research Program under BRSI-98-2418, and KOSEF 1998 Interdisciplinary Research Grant
98-07-02-07-01-5. Q-H. Park and H.J. Shin were supported in part by the program of Basic Science
Research, Ministry of Education 1998-015-D00073, and by Korea Science and Engineering Foundation,
97-07-02-02-01-3. K. Lee acknowledge an  useful discussion  with Karen Uhlenbeck about the
solutions of the chiral models, and K. Lee and Q-H. Park thank Aspen Center for Physics (1999
summer) where this work has been completed.

\end{document}